\documentclass[twocolumn]{aastex63}
\usepackage{amsmath}
\usepackage{booktabs}
\usepackage{comment}
\usepackage{xspace}
\usepackage{enumitem}
\usepackage{hyperref}
\usepackage{chronosys}
\usepackage{longtable}
\usepackage{latexsym}
\usepackage{amssymb}
\usepackage{longtable}
\usepackage{graphicx}
\usepackage{epsf}
\usepackage{url}

\received{}
\revised{}
\accepted{}

\shorttitle{\src AMXP Discovery}
\shortauthors{Ng et al.}

\newcommand{\src}{SRGA~J144459.2$-$604207\xspace}

\graphicspath{{./}{figures/}}

\begin{document}

\title{NICER Discovery that SRGA~J144459.2$-$604207 is an Accreting Millisecond X-ray Pulsar}

\correspondingauthor{Mason Ng}
\email{masonng@mit.edu}

\author[0000-0002-0940-6563]{Mason Ng}
\affiliation{MIT Kavli Institute for Astrophysics and Space Research, Massachusetts Institute of Technology, Cambridge, MA 02139, USA}

\author[0000-0002-5297-5278]{Paul S. Ray}
\affiliation{Space Science Division, U.S. Naval Research Laboratory, Washington, DC 20375, USA}

\author[0000-0002-0118-2649]{Andrea Sanna}
\affiliation{Dipartimento di Fisica, Universit\`a degli Studi di Cagliari, SP Monserrato-Sestu km 0.7, I-09042 Monserrato, Italy}

\author[0000-0001-7681-5845]{Tod E. Strohmayer}
\affiliation{Astrophysics Science Division, NASA Goddard Space Flight Center, Greenbelt, MD 20771, USA}
\affil{Joint Space-Science Institute, NASA Goddard Space Flight Center, Greenbelt, MD 20771, USA}

\author[0000-0001-6289-7413]{Alessandro Papitto}
\affiliation{INAF Osservatorio Astronomico di Roma, Via Frascati 33, I-00040 Monte Porzio Catone (RM), Italy}

\author[0000-0003-4795-7072]{Giulia Illiano}
\affiliation{INAF Osservatorio Astronomico di Roma, Via Frascati 33, I-00040 Monte Porzio Catone (RM), Italy}
\affiliation{Dipartimento di Fisica, Universit\`a degli Studi di Roma ``Tor Vergata," Via della Ricerca Scientifica 1, I-00133 Roma, Italy}
\affiliation{Dipartimento di Fisica, Universit\`a degli Studi di Roma ``La Sapienza," Piazzale Aldo Moro 5, I-00185 Roma, Italy}

\author[0000-0001-5472-0554]{Arianna C. Albayati}
\affiliation{School of Physics and Astronomy, University of Southampton, Southampton SO17 1BJ, UK}

\author[0000-0002-3422-0074]{Diego Altamirano}
\affiliation{School of Physics and Astronomy, University of Southampton, Southampton SO17 1BJ, UK}

\author[0000-0002-4729-1592]{Tu\u{g}ba Boztepe}
\affiliation{Istanbul University, Graduate School of Sciences, Department of Astronomy and Space Sciences, Beyaz\i t, 34119, \.Istanbul, T\"urkiye}

\author[0000-0002-3531-9842]{Tolga G\"uver}
\affiliation{Istanbul University, Science Faculty, Department of Astronomy and Space Sciences, Beyaz\i t, 34119, \.Istanbul, T\"urkiye}
\affiliation{Istanbul University Observatory Research and Application Center, Istanbul University 34119, \.Istanbul, T\"urkiye}

\author[0000-0001-8804-8946]{Deepto Chakrabarty}
\affiliation{MIT Kavli Institute for Astrophysics and Space Research, Massachusetts Institute of Technology, Cambridge, MA 02139, USA}

\author[0009-0008-6187-8753]{Zaven Arzoumanian}
\affiliation{Astrophysics Science Division, NASA Goddard Space Flight Center, Greenbelt, MD 20771, USA}

\author[0000-0002-5341-6929]{D. J. K. Buisson}\affiliation{Independent Researcher}

\author[0000-0001-7828-7708]{Elizabeth C. Ferrara}
\affiliation{Astrophysics Science Division, NASA Goddard Space Flight Center, Greenbelt, MD 20771, USA}
\affiliation{Department of Astronomy, University of Maryland, College Park, MD 20742, USA}
\affiliation{Center for Research and Exploration in Space Science \& Technology II (CRESST II), NASA/GSFC, Greenbelt, MD 20771, USA}

\author[0000-0001-7115-2819]{Keith C. Gendreau}
\affiliation{Astrophysics Science Division, NASA Goddard Space Flight Center, Greenbelt, MD 20771, USA}

\author[0000-0002-6449-106X]{Sebastien~Guillot}
\affil{IRAP, CNRS, 9 avenue du Colonel Roche, BP 44346, F-31028 Toulouse Cedex 4, France}
\affil{Universit\'{e} de Toulouse, CNES, UPS-OMP, F-31028 Toulouse, France.}

\author[0000-0002-8548-482X]{Jeremy Hare}
\affiliation{Astrophysics Science Division, NASA Goddard Space Flight Center, Greenbelt, MD 20771, USA}
\affiliation{Center for Research and Exploration in Space Science \& Technology II (CRESST II), NASA/GSFC, Greenbelt, MD 20771, USA}
\affiliation{The Catholic University of America, 620 Michigan Ave., N.E. Washington, DC 20064, USA}

\author[0000-0002-6789-2723]{Gaurava K. Jaisawal}
\affiliation{DTU Space, Technical University of Denmark, Elektrovej 327-328, DK-2800 Lyngby, Denmark}

\author[0000-0002-0380-0041]{Christian Malacaria} 
\affiliation{International Space Science Institute, Hallerstrasse 6, 3012 Bern, Switzerland}

\author[0000-0002-4013-5650]{Michael T. Wolff}
\affiliation{Space Science Division, U.S. Naval Research Laboratory, Washington, DC 20375, USA}

\begin{abstract}

We present the discovery, with the Neutron Star Interior Composition Explorer (NICER), that \src{} is a $447.9{\rm\,Hz}$ accreting millisecond X-ray pulsar (AMXP), which underwent a four-week long outburst starting on 2024 February 15. The AMXP resides in a 5.22~hr binary, orbiting a low-mass companion donor with $M_d>0.1M_\odot$. We report on the temporal and spectral properties from NICER observations during the early days of the outburst, from 2024 February 21 through 2024 February 23, during which NICER also detected a type-I X-ray burst that exhibited a plateau lasting $\sim 6$~s. The spectra of the persistent emission were well described by an absorbed thermal blackbody and power-law model, with blackbody temperature $kT\approx0.9{\rm\,keV}$ and power-law photon index $\Gamma\approx1.9$. Time-resolved burst spectroscopy confirmed the thermonuclear nature of the burst, where an additional blackbody component reached a maximum temperature of nearly $kT\approx3{\rm\,keV}$ at the peak of the burst. We discuss the nature of the companion as well as the type-I X-ray burst.


\end{abstract}


\section{Introduction} \label{sec:intro}

Accreting millisecond X-ray pulsars (AMXPs) are weakly magnetized ($B\sim10^8{\rm\,G}$) neutron stars with millisecond spin periods accreting from a low-mass companion donor, $M_d<1M_\odot$ \citep[see][for recent reviews]{disalvo22,disalvo23}. AMXPs are believed to have formed from the sustained spin-up torque provided by accreted material from a donor undergoing Roche lobe overflow over Gyr timescales; this is known as the recycling scenario \citep{alpar92}. AMXPs represent the most extreme spin periods for collapsed stellar objects, and thus are excellent laboratories for testing the limits of accretion physics \citep{chakrabarty03,bhattacharyya17}. By comparing the expected accretion luminosity produced by different orbital angular momentum sinks (e.g., magnetic braking and/or gravitational radiation) and the observed accretion luminosity during outbursts, AMXPs can be used to test mass transfer scenarios in these extreme systems \citep[e.g.,][]{marino19a,ngmason21,bult21a}. For example, in the latest outburst of SAX~J1808.4$-$3658, the observed orbital evolution was proposed to be due to ejected material having a specific angular momentum equal to or greater than that of the companion \citep{applegate94,illiano23}. We can further refine the physics governing these systems through the discovery of additional AMXPs with all-sky multi-wavelength monitoring of any transient outburst activity. These intermittent outbursts are likely the result of thermal instabilities in the accretion disk around the central neutron star \citep{lasota01}. 

The Mikhail Pavlinsky ART-XC telescope on the Spectrum-Roentgen-Gamma (SRG) observatory \citep{pavlinsky21,sunyaev21} first reported the discovery of a bright X-ray transient, \src, on 2024 February 21 \citep{atel16464}. Additional data from the MAXI/GSC X-ray sky monitor showed that the outburst actually began around February 15 \citep{atel16469}. This discovery was promptly followed up by an array of telescopes and instruments across many wavelengths. NICER discovered coherent X-ray pulsations at around $447.9{\rm\,Hz}$ and a type-I X-ray burst in its initial observations \citep{atel16474}. MeerKAT reported a non-detection at GHz radio wavelengths \citep{atel16475}, but the X-ray source localization provided by Chandra \citep{atel16510} led to the detection of a GHz radio counterpart by ATCA \citep{atel16511}. Optical follow-up observations did not detect an optical counterpart \citep{atel16476,atel16477,atel16487,atel16489}. A candidate near-infrared counterpart was identified with the PRIME telescope \citep{atel16499}, but it is formally incompatible with the Chandra localization. Further observations by INTEGRAL \citep{atel16485,atel16507} and NinjaSat \citep{atel16495} detected type-I X-ray bursts with recurrence timescales ranging from 1.7--2.9~h, with the period increasing as the persistent emission flux decreased.

We present the NICER discovery of 447.9~Hz pulsations of the AMXP \src and provide a preliminary timing solution for the orbital modulation of the pulse frequency. We also report on the discovery and characterization of the type-I X-ray burst. It is the 27th known AMXP and the first AMXP outburst discovered by the ART-XC telescope. We note that after this manuscript was submitted, the ART-XC team reported on ART-XC observations of the source \citep{molkov24}. In \S\ref{sec:observations}, we present the NICER X-ray observations undertaken and the associated analysis procedures. We present the timing and spectroscopic results in \S\ref{sec:results} and discuss the results in \S\ref{sec:discussion}.

\section{Observations and Data Analysis} \label{sec:observations}

The Neutron Star Interior Composition Explorer (NICER) is an external payload on the International Space Station. It consists of 56 (52 operational) co-aligned X-ray concentrator optics and silicon drift detectors in focal plane modules (FPMs). NICER provides fast-timing capabilities in the 0.2--12.0~keV energy range, and the onboard global positioning system (GPS) receiver allows for 100~ns time-tagging accuracy \citep{gendreau16,lamarr16,prigozhin16}.

We report on public NICER observations conducted over 2024 February 21 through 2024 February 23 (MJD 60361--60363), with observation IDs (ObsIDs) 6204190101--6204190103. Additional NICER observations were conducted under the NICER Guest Observer program (PI: A. Papitto) and will be reported elsewhere. Our observations were reduced and processed with \textsc{HEASoft} version 6.33 and the NICER Data Analysis Software (\textsc{NICERDAS}) version 12 (2024-02-09\_V012) using calibration version \texttt{xti20240206}. We imposed the following filtering criteria for the observations: angular offset for the source of $\texttt{ANG\_DIST}<54\arcsec$; NICER being outside of the South Atlantic Anomaly; an Earth limb elevation angle of $\texttt{ELV}>20^\circ$; a bright Earth limb angle of $\texttt{BR\_EARTH}>30^\circ$; an undershoot rate (per FPM; for dark current) of $\texttt{underonly\_range}=$ 0--500, and an overshoot rate (per FPM; for charged particle saturation) of $\texttt{overonly\_range}=$ 0--30. This resulted in good time intervals (GTIs) totaling 9.15 ks for scientific analysis.

We transformed the photon arrival times into the inertial frame of the Solar System by performing the Solar System barycenter corrections in the ICRS reference frame with source coordinates $\rm{R.A.}=221\fdg24558, {\rm\, Decl.}=-60\fdg69869$ \citep{atel16510} obtained by Chandra, using \texttt{barycorr} in FTOOLS with the JPL DE421 Solar System ephemeris \citep{folkner09}. We made use of \texttt{XSPEC} 12.14.0 \citep{arnaud96} for our spectral analysis. The generation of spectral products was enabled by the \texttt{nicerl3-spect} spectral product pipeline, which allowed us to: group the spectra with the optimal binning; rebin the spectra so that each bin had a minimum of 25 counts \citep{kaastra16}; generate background spectra using the \texttt{nibackgen3C50} model \citep{remillard22}, and generate the associated response matrices.

\section{Results} \label{sec:results}

\begin{figure*}[htbp!]
    \centering
    \includegraphics[width=1.05\textwidth]{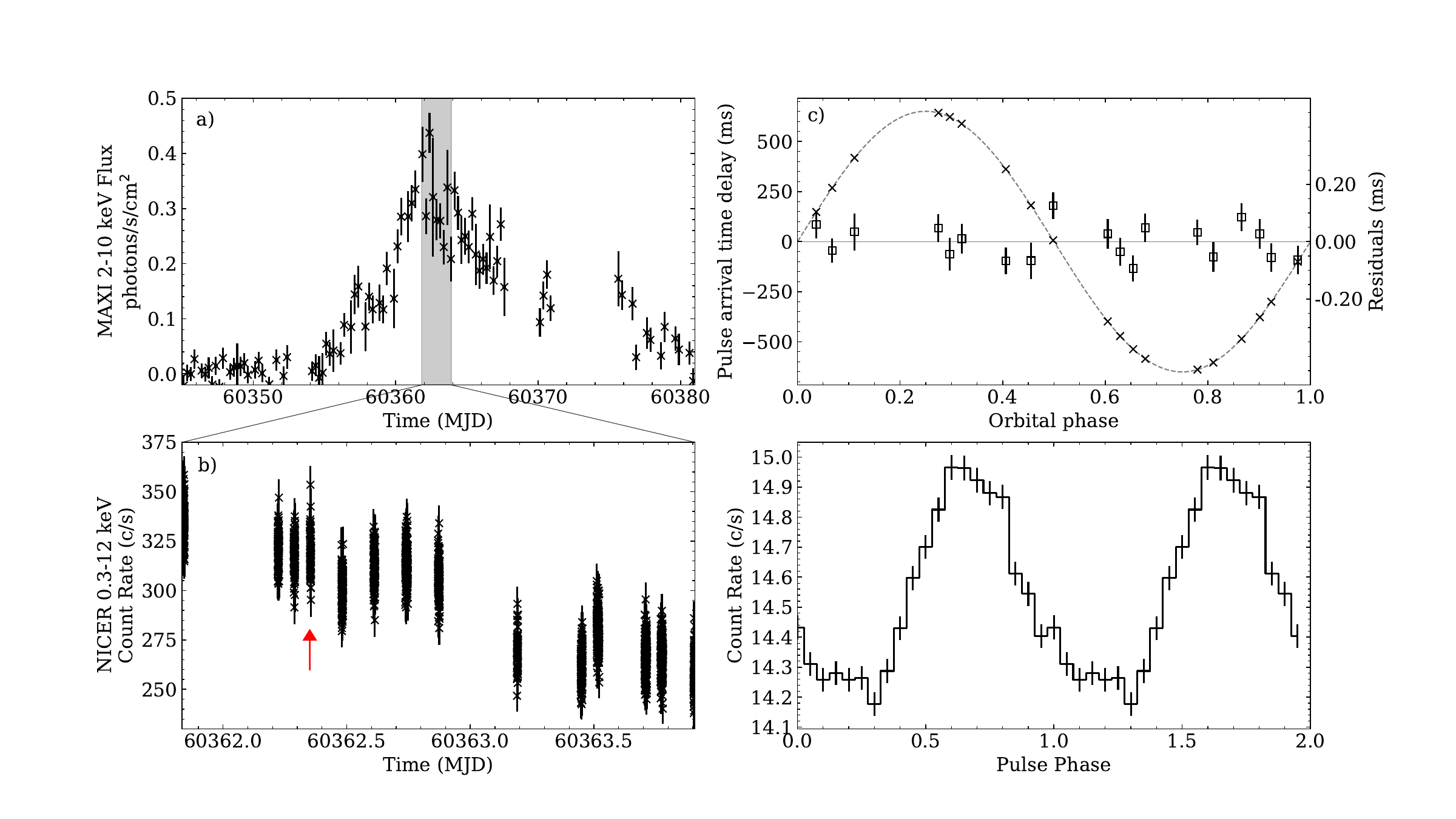}
    \caption{a) \src monitoring 2--10~keV light curve by MAXI/GSC \citep{atel16469}. The shaded gray region corresponds to the NICER observation interval. b) NICER 0.3--12.0~keV light curve with 4~s bins. The burst emission was removed, but instead the midpoint of the burst interval (including the pre-burst) is shown with the red arrow. c) The pulse arrival time delay as a function of the orbital phase, with the best-fit orbit shown with the dashed lines, and the pulse timing residuals shown in squares (with scale on the right y-axis). d) Folded 1.0--10.0~keV pulse profile with the NICER observations. The profile also shows a significant second harmonic. There are 20 pulse phase bins and two cycles are plotted for clarity.}
    \label{fig:lc_phase}
\end{figure*}

We place the NICER observations in context of the overall outburst evolution from MAXI/GSC observations in Figures~\ref{fig:lc_phase}(a) and (b) -- as reported in the figure zoom-in, the NICER observations analyzed were around the peak of the outburst.

\subsection{X-ray Pulsation Discovery and Timing} \label{sec:timing}

To search for coherent periodicity, we employed the Fourier-domain acceleration search (FDAS) method implemented in version 4.0 of the open-source pulsar search and analysis software \texttt{PRESTO} \citep{ransom11}. The algorithm accounts for possible pulsation frequency modulations due to the binary Doppler effect with orbital period $P_{\rm orb}$ \citep{ransom02}. The FDAS were carried out on individual GTIs throughout the observation span of \src, with a minimum interval length of $100{\rm\,s}$. We had 15 GTIs overall, with a median segment length of 613~s (spanning 183~s to 911~s). We adopted this approach as the sensitivity of acceleration searches is optimized for segment lengths $T$ such that $T\lesssim P_{\rm orb}/10$, as the pulsar acceleration is approximately constant and the spin frequency is expected to linearly evolve within the segment \citep{ransom02}. We searched over the 1--1000~Hz frequency range, and in two energy bands: 0.3--2.0~keV and 2.0--10.0~keV. We found significant candidates in several, but not all, GTIs within the 2.0--10.0~keV band. Specifically, we found ten candidates around 447.9~Hz with single-trial significance over 2.5--5.5$\sigma$. Given the overall pulsation search parameter space, the pulsation signal corresponds to a trials-adjusted significance of $11.3\sigma$ ($5.4\times10^{10}$ trials) across all of the independent data segments, frequency bins, and Fourier bins searched.

With the secure detection of coherent X-ray pulsations, we proceeded to characterize the pulse. First, we derived a provisional circular orbit model by fitting an ellipse in the period-acceleration plane with the ten significant candidates. Next, we performed an epoch folding search in approximately 1~ks intervals and modeled the residual frequency variations, resulting in revised orbital parameters from the initial estimate. We also maximized the profile variance ($\chi^2$) from exploring a grid of values defined by the spin frequency ($\nu_0$) and epoch of ascending node passage ($T_{\rm asc}$). Next, we optimized the pulse significance by calculating the $H$-statistic \citep{dejager89} with two harmonics ($m=2$) 
\begin{equation}
    H \equiv \underset{1\leq m \leq2}{\rm max}(Z_m^2-4m+4),
\end{equation}
for 
\begin{equation}    
    Z_m^2 = \frac{2}{N}\left[\left(\sum\limits_{j=1}^N \text{cos}\,2m\pi\nu t_j\right)^2 + \left(\sum\limits_{j=1}^N\text{sin}\,2m\pi\nu t_j\right)^2\right],
\end{equation}
where $t_j$ are the photon arrival times corrected for Doppler modulation described by the provisional orbit model ($j\in\{1,\ldots,N\}$ for $N$ photons), over a grid of energy values \citep{buccheri83}. We found that the $H$-statistic was maximized for the energy range 1.03--11.97~keV. However, the $H$-statistic exhibited very little variance (maximum of $\Delta H=+2$) throughout the energy grid search. We adopted an energy range of 1.0-10.0~keV for the timing analysis as the background (see below) dominates below and above this energy range.

With this initial model, we generated 19 pulse times of arrival (TOAs) with the \texttt{photon\_toa.py} tool in the \texttt{NICERsoft} data analysis package\footnote{\url{https://github.com/paulray/NICERsoft/}}, where each TOA had an integration time of 300~s (with minimum exposure time of 200~s). The TOAs were then correlated with a pulse template, using all of the data, comprising the sum of three Gaussians. Finally, we fit the pulse TOAs to the \texttt{ELL1} binary orbit model \citep{lange01} available within PINT, an open-source pulsar timing \texttt{Python} package \citep{luo21}. The best-fit orbital ephemeris is given in Table~\ref{tab:ephemeris}, and the corresponding pulse arrival time delay (with respect to a constant spin frequency model) illustrating the best-fit orbit solution is shown in Figure~\ref{fig:lc_phase}(c). We note that the final orbital solution presented here is significantly different than the preliminary solutions previously reported both with the same NICER data \citep{atel16480} and from Insight-HXMT \citep{atel16548}. However, of these three solutions, the final solution we present here has a significantly larger maximum $H$-statistic value of 867.61, compared to 149.78 and 60.43, respectively. The final folded pulse profile is shown in Figure~\ref{fig:lc_phase}(d), where fitting a two-component sinusoid yielded fractional root-mean-squared amplitudes of $1.79\pm0.04\%$ and $0.35\pm0.04\%$, for the fundamental and second harmonic, respectively. The evolution of the pulse profile throughout the outburst and its energy dependence is outside the scope of this Letter, and will be reported elsewhere.

\begin{table*}[t]
\centering
\caption{
  Timing model for \src from the 2024 February outburst.
  \label{tab:ephemeris}
}  
\begin{tabular}{lc}
\toprule 
\toprule
Parameter & Value \\
\toprule 
Right Ascension, $\alpha$ (J2000) & $221\fdg2455833$ \\
Declination, $\delta$ (J2000) & $-60\fdg6986944$ \\
Spin frequency, $\nu_0$ (Hz) & 447.87156100(11) \\
Spin epoch, $t_0$ (TDB) & MJD $60362.87145091$ \\ 
Binary period, $P_{\rm orb}$ (d) & 0.2176354(5) \\
Projected semimajor axis, $a_x \sin i$ (lt-s) & 0.650527(17) \\
Epoch of ascending node passage, $T_{\rm asc}$ (TDB) & MJD 60361.858933(3)  \\ 
Eccentricity, $e$ & $<4\times10^{-4}\ (3\sigma$) \\
$\chi^2/{\rm d.o.f.}$ & 27.7/11 \\
\bottomrule
\end{tabular}

\tablecomments{The Solar System barycenter corrections were performed using the source coordinates determined with Chandra High Resolution Camera observations of the source region \citep{atel16510}.}
\end{table*}

\subsection{Type-I X-ray Burst} \label{sec:type1burst}

\begin{figure}
    \centering
    \includegraphics[width=0.5\textwidth,height=\textheight,keepaspectratio]{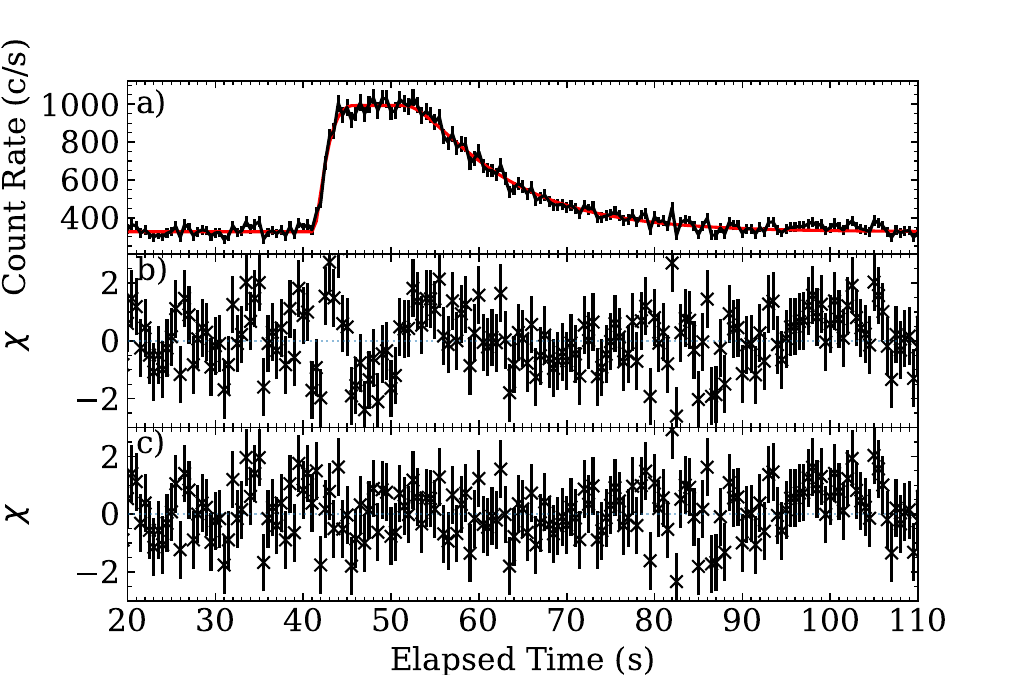}
    \caption{Phenomenological fit to the light curve of the type-I X-ray burst in the NICER ObsID 6204190102 with a simple FRED model and a model with a FRED component and a plateau. a) Light curve of the type-I X-ray burst, with 0.5~s bins, with the FRED + plateau model shown in the solid red line; b) Residuals from fitting the burst light curve to a simple FRED model; c) Residuals from a FRED + plateau model. While the fit with the FRED model is formally acceptable ($\chi^2/{\rm d.o.f.}=1.08$; 318 d.o.f.), the fit with the hybrid FRED and plateau model is significantly better at the $1.4\times10^{-11}$ level, and the systematic residuals around the plateau are minimized. The elapsed time of 0 corresponds to MJD 60362.35007 (TT units at NICER).}
    \label{fig:burst_fit}
\end{figure}

During the NICER observations, we detected a sharp, short-lived increase in the X-ray flux, which the light curve evolution suggested was a type-I X-ray burst. We first characterized the light curve ($f(t)$ as a function of time, $t$) by fitting with a hybrid fast-rise exponential-decay (FRED) model with a plateau \citep{barriere15}, described by
\begin{equation}
    f(t) = \begin{cases} C & t \leq t_0 \\ 
    A\,\text{exp}\left(-\frac{\tau_R}{t-t_0} - \frac{t-t_0}{\tau_D}\right) + C & t_0 < t \leq t_{\rm pe} \\ 
    A\,\text{exp}\left(-2\sqrt{\frac{\tau_R}{\tau_D}}\right) + C & t_{\rm pe} < t \leq t_{\rm pe} + t_{\rm pl} \\ 
    A\,\text{exp}\left( -\frac{\tau_R}{t-t_0-t_{\rm pl}} - \frac{t-t_0-t_{\rm pl}}{\tau_D}\right) + C & t > t_{\rm pe} + t_{\rm pl}
    \end{cases}
\end{equation}
where $C$ is the persistent flux level (c/s), $A$ is the burst amplitude (c/s), $t_0$ is the burst onset time (in seconds), $\tau_R$ and $\tau_D$ are the rise and decay timescales (in seconds), respectively, $t_{\rm pe} = t_0 + \sqrt{\tau_R\tau_D}$ is the time at the peak (in seconds), and $t_{\rm pl}$ is the interval length of the plateau (in seconds). The corresponding best fit parameters are given in Table~\ref{tab:burst_table}. We show the results of the fit in Figure~\ref{fig:burst_fit}, where we also show residuals from fitting a simple FRED model (Figure~\ref{fig:burst_fit}b). While both models provide a reasonable description of the burst, the hybrid FRED and plateau model significantly improves the residuals around the burst peak. The simple FRED model can also be rejected with probability $1.4\times10^{-11}$ compared to the hybrid FRED and plateau model according to the Akaike information criterion \citep{akaike74,liddle07}.

\begin{deluxetable}{ccc}[htbp]
\caption{Best-fit parameters from fitting a FRED or a FRED + plateau model to the type I X-ray burst light curve with a reference time of MJD 60362.350531(2) (TT units at NICER).}
\tablehead{
\colhead{Parameter} & \colhead{FRED} & \colhead{FRED + Plateau}}  
\startdata
$t_0$ (s) & $39.1\pm0.3$ & $40.7\pm0.2$ \\ 
$t_{\rm pl}$ (s) & \nodata & $5.8\pm0.7$ \\ 
$\tau_D$ (s) & $10.0\pm0.3$ & $9.3\pm0.4$ \\
$\tau_R$ (s) & $7.7\pm1.0$ & $2.7\pm0.5$ \\ 
$A$ (c/s) & $4300\pm600$ & $1900\pm200$ \\ 
$C$ (c/s) & $324.3\pm1.8$ & $326.0\pm1.7$ \\ 
\midrule
$\chi^2/{\rm d.o.f.}$ & $343/318$ & $295/317$
\enddata
\tablecomments{The uncertainties are given to $1\sigma$ confidence level.}
\label{tab:burst_table}   
\end{deluxetable}

\subsection{Spectroscopy} \label{sec:spectroscopy}

\subsubsection{Persistent Emission}

First, we defined several intervals for spectral extraction. The pre-burst interval was between MJDs 60362.35006 to 60362.35047 (TT units at NICER; spanning about 35~s), 5~s before the burst onset. The burst interval was defined to be between MJDs 60362.35047 to 60362.35146 (spanning about 85~s), 80~s after the burst onset. We also present fits to the individual ObsIDs, but for ObsID 6204190102 which contains the type-I X-ray burst, we excised the pre-burst and burst intervals.

For the spectral fits, we adopted a line of sight hydrogen column density of $n_H = 2.9\times10^{22}{\rm\,cm^{-2}}$. This value was determined from a joint fit of the spectra from the three ObsIDs (not shown; pre-burst and burst intervals removed), where we employed an absorbed thermal blackbody and power-law model and untied all spectral parameters except for $n_H$, and we found $n_H = 2.90\pm0.03\times10^{22}{\rm\,cm^{-2}}$. All subsequent spectral fitting was restricted to 1.0--10.0~keV as the soft X-rays were absorbed below around 1.0~keV because of the high $n_H$, and the background dominated above 10~keV. The \texttt{wilm} elemental abundance model was adopted for the spectral fits \citep{wilms00}. During the course of the three reported observations, the blackbody normalization remained constant within uncertainties, with an average value of ${\rm norm_{\rm BB}} = 36.2_{-5.8}^{+7.4}\ (R_{\rm km}/D_{10})^2$, where $R_{\rm km}$ is the source region radius in km, and $D_{10}$ is the source distance in units of 10~kpc. The blackbody temperature marginally decreased from $kT = 0.99\pm0.08{\rm\,keV}$ to $kT = 0.85_{-0.06}^{+0.07}{\rm\,keV}$, the photon index increased from $\Gamma=1.81\pm0.03$ to $\Gamma=1.90\pm0.02$, and the power-law normalization marginally decreased from ${\rm norm_{\rm PL}} = 0.76\pm0.03{\rm\,photons/s/cm^2/keV}$ to ${\rm norm_{\rm PL}} = 0.69\pm0.03{\rm\,photons/s/cm^2/keV}$. The results are reported in Table~\ref{tab:spec_table}. All uncertainties are reported at $90\%$ confidence levels. The detailed spectroscopic evolution of the persistent emission during the entire outburst will be presented in a separate publication.

\begin{deluxetable}{cccc}[htbp]
\caption{Spectroscopic results for individual ObsIDs with an absorbed thermal blackbody and power-law (\texttt{tbabs(bbodyrad+powerlaw)} in \texttt{XSPEC}).}
\tablehead{
\colhead{} & \multicolumn{3}{c}{\uline{ObsID}} \\ 
\colhead{Parameter} & \colhead{$6204190101$} & \colhead{$6204190102$} & \colhead{$6204190103$}}  
\startdata
\midrule
kT & $0.99_{-0.08}^{+0.08}$ & $0.95_{-0.06}^{+0.07}$ & $0.85_{-0.06}^{+0.07}$ \\
${\rm norm}_{\rm BB}$ & $38_{-10}^{+13}$ & $35_{-9}^{+11}$ & $35_{-11}^{+15}$ \\ 
$\Gamma$ & $1.81\pm0.03$ & $1.82\pm0.02$ & $1.90\pm0.02$ \\ 
${\rm norm}_{\rm PL}$ & $0.76\pm0.03$ & $0.73\pm0.03$ & $0.69\pm0.03$ \\ 
$F_{\rm 1-10{\rm\,keV}}$ & $2.648_{-0.033}^{+0.014}$ & $2.448_{-0.013}^{+0.008}$ & $2.009_{-0.014}^{+0.006}$ \\ 
\midrule 
$\chi^2/{\rm d.o.f.}$ & 98.2/123 & 141.0/139 & 148.7/137 \\  
\enddata
\tablecomments{The uncertainties are given to $90\%$ confidence level. The parameter ${\rm norm}_{\rm BB}$ is scaled by ($R_{\rm km}/D_{10})^2$ and ${\rm norm}_{\rm PL}$ has units of ${\rm photons/keV/cm^2/s}$. The absorbed 1.0--10.0~keV flux, $F_{\rm 1-10{\rm\,keV}}$ is expressed in units of $10^{-9}{\rm\,erg\,s^{-1}\,cm^{-2}}$.}
\label{tab:spec_table}   
\end{deluxetable}

Next, we analyzed the pre-burst and burst intervals. The pre-burst spectrum was well described by an absorbed thermal blackbody and power-law model (\texttt{tbabs(bbodyrad+powerlaw)}), where $n_H=2.9\times10^{22}{\rm\,cm^{-2}}$ (fixed as per above), $kT = 0.99_{-0.19}^{+0.20}{\rm\,keV}$, ${\rm norm}_{\rm BB} = 53_{-25}^{+57} (R_{\rm km}/D_{10})^2$, photon index $\Gamma = 1.92_{-0.10}^{+0.14}$, and ${\rm norm}_{\rm PL} = 0.71\pm0.07{\rm\,photons/s/cm^2/keV}$.

\subsubsection{Time-Resolved Burst Spectroscopy}

\begin{figure}
    \centering
    \includegraphics[width=0.5\textwidth,height=\textheight,keepaspectratio]{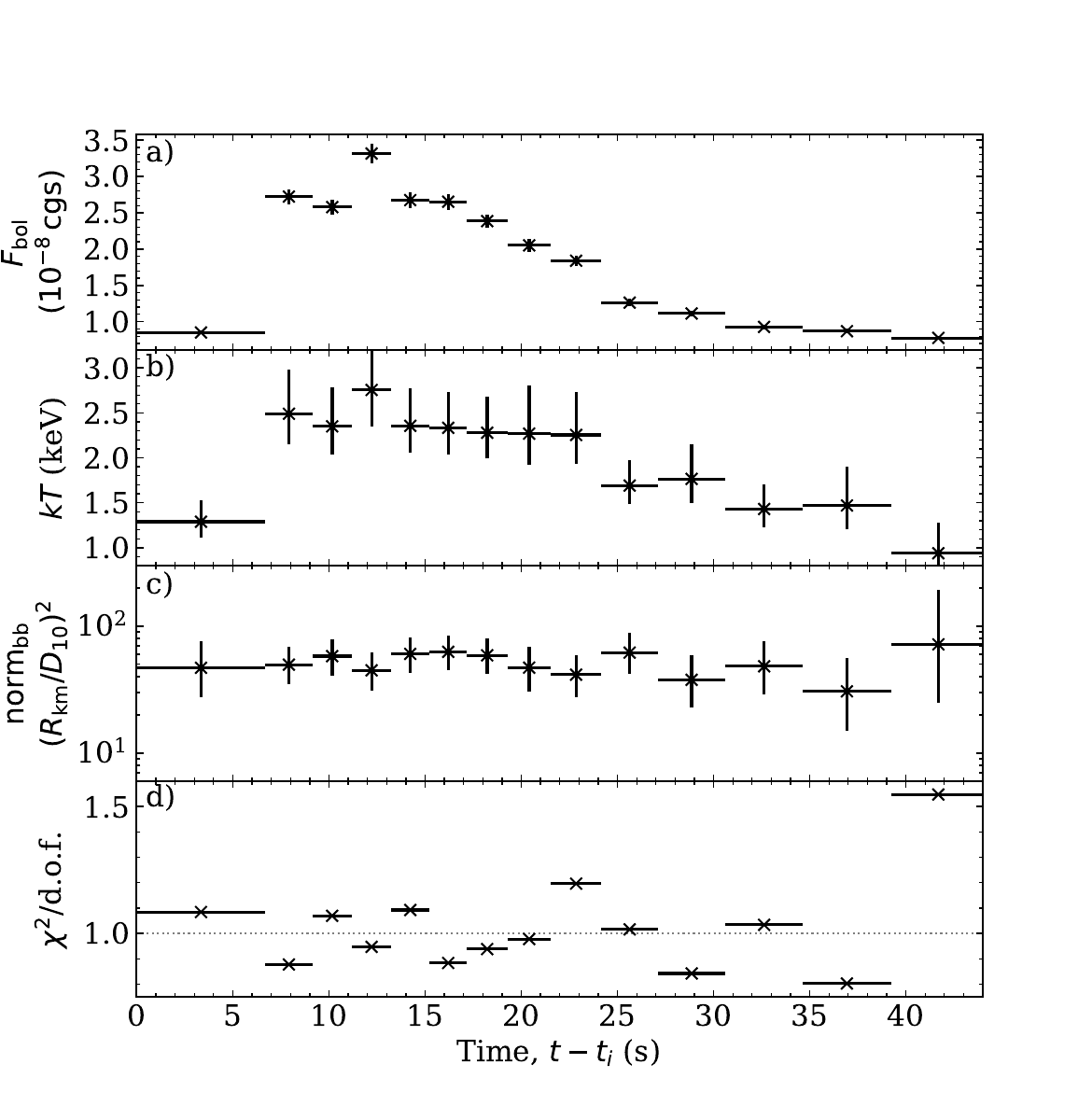}
    \caption{Evolution of the second blackbody component during the type-I X-ray burst from time-resolved spectroscopy. a) bolometric (0.1--100.0~keV) unabsorbed flux in units of $10^{-8}{\rm\,erg\,s^{-1}\,cm^{-2}}$; b) blackbody temperature, $kT$ (keV); c) blackbody normalization, ${\rm norm}_{\rm bb}$ scaled by $(R_{\rm km}/D_{10})^2$; d) reduced $\chi^2$ for the fit. The cooling blackbody, the key signature of the thermonuclear nature of the burst, is evident. The offset, $t_i$, corresponds to MJD~60362.35047 (TT units at NICER).}
    \label{fig:burst_spec}
\end{figure}

To further characterize the burst, we performed time-resolved spectroscopy starting from 5 s before burst onset, to 40 s after burst onset. We generated dynamically binned spectra, ensuring that each spectrum contained at least 2000 photons, and used \texttt{nicerl3-spect} to generate the associated responses. We used the \texttt{nibackgen3C50} background spectrum corresponding to ObsID 6204190102, as the background is not expected to vary significantly over the burst. We also regrouped the spectra such that each spectral bin had a minimum of 25 counts. Given that the background dominated above 6.5~keV, the spectral fits were restricted to 1.0--6.5~keV. We fixed the parameters of the pre-burst model, and added a second thermal blackbody model to account for the burst evolution (the full model is \texttt{tbabs(bbodyrad+powerlaw+bbodyrad)}), which is shown in Figure~\ref{fig:burst_spec}. The burst decay was well-described by a cooling thermal blackbody component, which supports the thermonuclear origin for the X-ray burst. Motivated by bursters which have exhibited excess emission above the burst, we tried to fit the spectra by including a scaling factor to the underlying pre-burst emission \citep{worpel13,worpel15,guver22a,guver22b}, but it did not statistically improve the spectral fits. This is similar to what has been seen in sources with relatively higher hydrogen column density values of $n_H > 10^{22}{\rm\,cm^{-2}}$ \citep{bult21b,guver21,bostanci23}. The bolometric unabsorbed flux shown in Figure~\ref{fig:burst_spec}(a) was derived using the convolutional \texttt{cflux} model, and we extrapolated the model out to 0.1--100~keV by using the \texttt{energies} command in \texttt{XSPEC}.

\section{Discussion} \label{sec:discussion}

We have presented the NICER discovery of $447.9{\rm\,Hz}$ pulsations of the accreting millisecond X-ray pulsar \src, which resides in a 5.22~hr binary. We reported on initial NICER observations of the source around the peak of the outburst, from 2024 February 21 to 2024 February 23. The observations also revealed a type-I X-ray burst, and time-resolved burst spectroscopy confirmed the thermonuclear nature of the burst. 

\begin{figure}
    \centering
    \includegraphics[width=0.5\textwidth,height=\textheight,keepaspectratio]{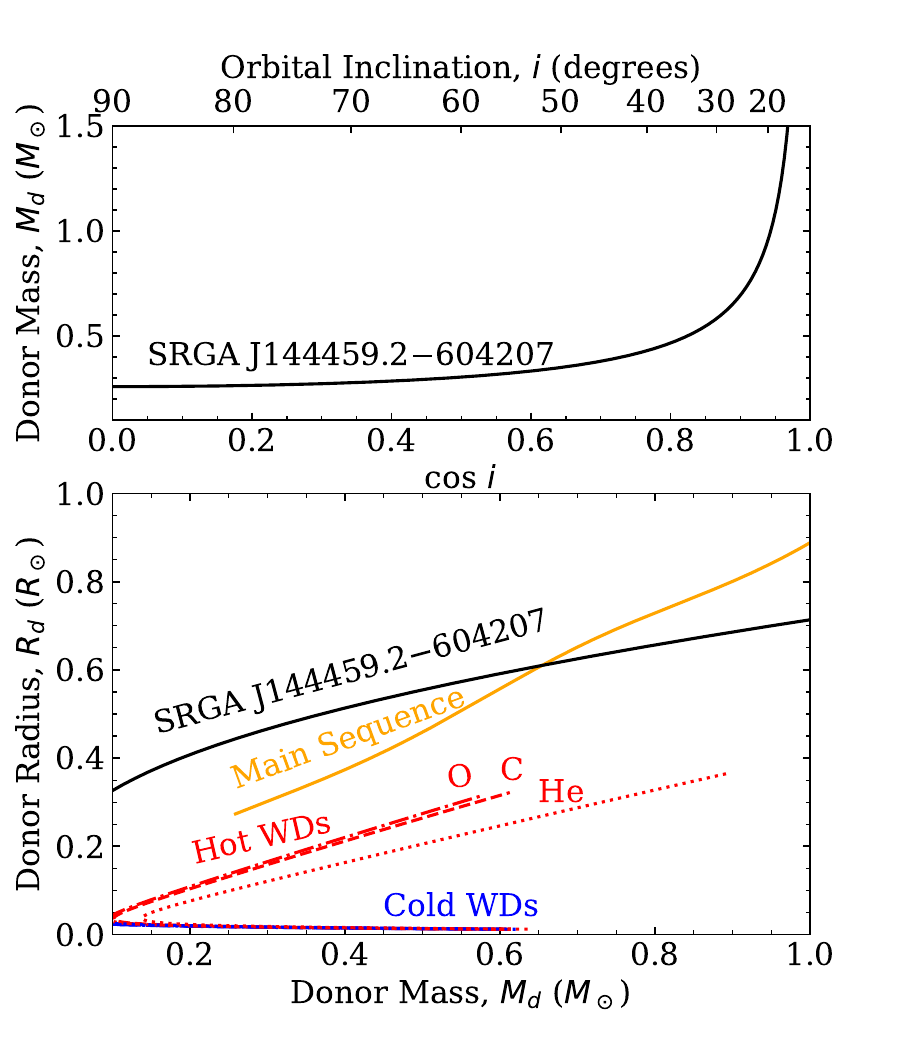}
    \caption{Top: Donor mass ($M_d$, $M_\odot$) as a function of the binary inclination, $i$, for $M_{\rm ns}=1.4M_\odot$; $1-\text{cos}\,i$ is the a priori cumulative distribution function of observed binary inclinations. Bottom: Donor radius ($R_d$, $R_\odot$) as a function of the donor mass for \src (black solid line), zero-age main sequence stars \citep[orange solid line;][]{tout96}, and hot (red) and cold (blue) WDs of varying compositions \citep{deloye03}. The most viable donor for \src is a main sequence star (which may be slightly bloated).}
    \label{fig:donorMR}
\end{figure}

The determination of the binary orbital parameters of the \src system allows us to constrain the properties of the companion/mass donor. We can employ the binary mass function, 
\begin{equation} \label{eq:massfunction}
    f_m = \frac{(M_d\text{sin}\,i)^3}{(M_{\rm ns} + M_d)^2} = \frac{4\pi^2 (a_x\text{sin}\,i)^3}{GP_{\rm orb}^2},
\end{equation}
where $M_d$ and $M_{\rm ns}$ are the donor and neutron star masses, respectively, and $G$ is the gravitational constant. Assuming $M_{\rm ns}=1.4 M_\odot$, we plot the donor mass-radius curve from Equation~\ref{eq:massfunction} assuming a Roche lobe-filling donor, which is shown in Figure~\ref{fig:donorMR}. We also plot mass-radius curves for several companion types, including white dwarfs \citep[WDs;][]{deloye03} and zero-age main sequence stars \citep{tout96}. We found that for all plausible cold (core temperature, $T_c<10^6{\rm\,K}$) and hot ($T_c>10^6{\rm\,K}$) WDs, the WDs are too small to fill their Roche lobes to be viable companions for \src.

On the other hand, main sequence (H-rich) companions are much more likely. For semi-detached binaries such as the Roche lobe-filling donor and AMXP here, the period-density relation $\bar\rho \simeq 107(P_{\rm orb}/{1{\rm\,hr}})^{-2}{\rm\,g\,cm^{-3}}$ \citep{knigge11}
estimates a stellar mean density, $\bar{\rho}$ of around $\bar\rho\approx4.0{\rm\,g\,cm^{-3}}$, which suggests an early M-type main sequence donor \citep{drilling00}. While the mass-radius curve presented here is for solar metallicity ($Z=Z_\odot=0.02$), for donor masses below $1\,M_\odot$, the mass-radius curves are weakly sensitive to the metallicity \citep{tout96}. The mass-radius curves for \src and the main sequence companion intersect at around $0.65\,M_\odot$, which corresponds to an a priori probability over binary inclination of $P(i\leq i_{\rm 0.65\,M_\odot})\approx11\%$ assuming an isotropic population of binary inclinations. Consequently, most of the probability space, after considering binary inclination, lies above the main sequence in the mass-radius plane, indicating stars larger than a main sequence star at the same mass. The main sequence companion could have a bloated H atmosphere due to the X-ray irradiation by the neutron star, thus allowing for less massive (i.e., more probable) donors. Additionally, since we do not have full orbital coverage with this current data set, and given the non-contiguous nature of NICER observations, we cannot provide further inclination constraints through searches for dips or eclipses.

It is also much more plausible that the companion star is very old ($\sim$Gyr) given the recycled nature of the system \citep[e.g.,][]{altamirano08b,sanna18}, and the stellar isochrones for these old companions suggest a lower (more plausible) mass \citep{girardi00}. The spin and orbital properties of \src, along with the derived companion properties, fit well within the population of known AMXPs with main sequence companions \citep[e.g.,][]{papitto07,riggio11,altamirano11,papitto11b,papitto13,patruno21}. While we used $M_{\rm ns} = 1.4 M_\odot$, we note that the mass-radius relation for \src depends weakly on the NS mass. However, recycled pulsars which experienced low levels of sustained accretion over long timescales ($\sim$Gyr) should have higher masses \citep{romani90,romani22}.

NICER has detected a type-I X-ray burst within the observation interval covered in this Letter. The burst light curve profile was best fit with a phenomenological hybrid fast-rise exponential decay and plateau model (see Table~\ref{tab:burst_table}). The burst duration of $< 100{\rm\,s}$ suggests low accretion rates of mixed H/He material \citep{fujimoto87,intzand05}. Time-resolved burst spectroscopy as shown in Figure~\ref{fig:burst_spec} revealed a constant blackbody normalization during the burst, suggesting that a photospheric radius expansion (PRE) event did not take place. However, we can derive an upper limit on the source distance assuming that the burst peak was at Eddington luminosity. The peak bolometric (0.1--100.0~keV) unabsorbed flux was around $2.8\times10^{-8}{\rm\,erg\,s^{-1}\,cm^{-2}}$, and given the empirical critical luminosity of $3.8\times10^{38}{\rm\,erg\,s^{-1}}$ at the peak of type-I X-ray bursts during PRE events, this implies an upper limit on the source distance of $d<10.6{\rm\,kpc}$ \citep{kuulkers03}. 

Subsequent NICER observations, as well as those by other instruments such as Swift/XRT \citep{atel16475}, INTEGRAL \citep{atel16485,atel16507}, and NinjaSat \citep{atel16495} have detected more type-I X-ray bursts from \src. In fact, INTEGRAL reported the ``quasi-periodic" nature of the type-I X-ray burst trains, with a burst recurrence rate that is a function of the outburst flux \citep{atel16507}. Earlier INTEGRAL observations noted a recurrence timescale of about 1.7~h over 60~ks, implying a remarkably stable accretion rate \citep{atel16485}. There are three other known sources that have exhibited clocked bursting behavior. The canonical Clocked Burster (GS~1826$-$238) had regular burst recurrence timescales between 3.6--5.7~hr since its outburst onset in 1988 \citep{ubertini99,chenevez16}, though its bursting activity has dramatically reduced upon its transition into the spectrally softer ``banana'' state in 2016 \citep{chenevez16,yun23}. The other known clocked burster, 1RXS~J180408.9$-$342058, was found to display clocked bursting behavior in its intermediate spectral state, with a recurrence timescales of around 0.9~h \citep{marino19b}. The 11~Hz burster IGR~J17480$-$2446 also exhibited clock-like bursting activity \citep{chakraborty11}. Such behavior is thought to be related to the near-solar composition of the accreted material \citep{galloway04,heger07b,lampe16,meisel18}. We note that this is the second clocked burster for which we know the spin period and orbital properties, the other being IGR~J17480$-$2446 \citep{strohmayer10,papitto11a}. A detailed study of the broadband timing and spectral properties of the type-I X-ray bursts from the outburst of \src is outside of the scope of this Letter and will be reported in a future publication.

Assuming an averaged unabsorbed bolometric (0.1--100.0~keV) flux across the three observations of $F=7\times10^{-9}{\rm\,erg\,s^{-1}\,cm^{-2}}$, then for $d<10.6{\rm\,kpc}$, the source luminosity is estimated to be $L<9.5\times10^{37} {\rm\,erg\,s^{-1}}$. If we assume spin equilibrium, then the dipolar NS magnetic field is approximately 
\begin{equation}
    B = 4.2\zeta^{-\frac{7}{6}}\left(\frac{P}{10{\rm\,ms}}\right)^{\frac{7}{6}}\left(\frac{M_{\rm ns}}{1.4\ M_\odot}\right)^{\frac{1}{3}} \left(\frac{\dot{M}}{10^{-10}{\rm\,M_\odot/yr}}\right)^{\frac{1}{2}} 10^8\ G,
\end{equation}
where $\zeta$ is the order-unity ratio of the magnetospheric and Alfv\'en radii \citep{ghoshlamb79}, $P=1/\nu$ is the pulsar spin period, and $\dot{M}$ is the mass transfer rate from the donor onto the NS. For a NS with radius $R=10{\rm\,km}$ and $M_{\rm ns}=1.4M_\odot$, we find for $d<10.6{\rm\,kpc}$ that $\dot{M}<8.11\times10^{-9} (d/10.6{\rm\, kpc})^{2} {\rm\,M_\odot/yr}$. Thus for $d<10.6{\rm\,kpc}$, $B<(6.6-96.6)\times10^8{\rm\,G}$, which is within expectations (albeit a large range) from the distribution of estimated magnetic field strengths of AMXPs \citep{mukherjee15}. 

A more accurate determination of the magnetic field strength will come from observing more outbursts from \src to measure a long-term spin frequency derivative, $\dot\nu$. In fact, MAXI/GSC found signs of past weak activity in 2022 January and 2023 December \citep{atel16483}, and INTEGRAL reported significant hard X-ray activity in 2023 December \citep{atel16493}, but there were no dense monitoring observations available for those periods. Further X-ray monitoring observations will be crucial in detecting the rising outbursts of \src to further constrain the source properties, such as the evolution of the type-I X-ray burst morphology. Further all-sky high-cadence multiwavelength surveys will be instrumental in discovering more AMXPs to understand the population and diversity of phenomena they exhibit. 

\facilities{NICER, MAXI}

\software{Astropy \citep{astropy:2013, astropy:2018}, PRESTO \citep{ransom02}, NumPy and SciPy \citep{virtanen20}, Matplotlib \citep{hunter07}, IPython \citep{perez07}, tqdm \citep{dacostaluis22}, HEASoft 6.33\footnote{http://heasarc.gsfc.nasa.gov/ftools} \citep{heasoft}}

\acknowledgments

This research has made use of data and/or software provided by the High Energy Astrophysics Science Archive Research Center (HEASARC), which is a service of the Astrophysics Science Division at NASA/GSFC and the High Energy Astrophysics Division of the Smithsonian Astrophysical Observatory. This research has made use of MAXI data provided by RIKEN, JAXA and the MAXI team. NICER work at NRL is supported by NASA. M.N. was supported in part by NASA through the NICER Guest Observer Program. A.P. and G.I. acknowledge financial support from the National Institute for Astrophysics (INAF) Research Grant `Uncovering the optical beat of the fastest magnetised neutron stars (FANS)' and the Italian Ministry of University and Research (MUR) under PRIN 2020 grant No. 2020BRP57Z `Gravitational and Electromagnetic-wave Sources in the Universe with current and next-generation detectors (GEMS)'. T.G. and T.B. are supported by the Scientific Research Projects Coordination Unit of Istanbul University (ADEP Project No: FBA-2023-39409). J.H. acknowledges support from NASA under award number 80GSFC21M0002.

\bibliography{ulx}

\begin{thebibliography}{}
\expandafter\ifx\csname natexlab\endcsname\relax\def\natexlab#1{#1}\fi
\providecommand{\url}[1]{\href{#1}{#1}}
\providecommand{\dodoi}[1]{doi:~\href{http://doi.org/#1}{\nolinkurl{#1}}}
\providecommand{\doeprint}[1]{\href{http://ascl.net/#1}{\nolinkurl{http://ascl.net/#1}}}
\providecommand{\doarXiv}[1]{\href{https://arxiv.org/abs/#1}{\nolinkurl{https://arxiv.org/abs/#1}}}

\bibitem[{{Akaike}(1974)}]{akaike74}
{Akaike}, H. 1974, IEEE Transactions on Automatic Control, 19, 716

\bibitem[{{Alpar} {et~al.}(1992){Alpar}, {Hasinger}, {Shaham}, \&
  {Yancopoulos}}]{alpar92}
{Alpar}, M.~A., {Hasinger}, G., {Shaham}, J., \& {Yancopoulos}, S. 1992, \aap,
  257, 627

\bibitem[{{Altamirano} {et~al.}(2008){Altamirano}, {Casella}, {Patruno},
  {Wijnands}, \& {van der Klis}}]{altamirano08b}
{Altamirano}, D., {Casella}, P., {Patruno}, A., {Wijnands}, R., \& {van der
  Klis}, M. 2008, \apjl, 674, L45, \dodoi{10.1086/528983}

\bibitem[{{Altamirano} {et~al.}(2011){Altamirano}, {Cavecchi}, {Patruno},
  {Watts}, {Linares}, {Degenaar}, {Kalamkar}, {van der Klis}, {Rea}, {Casella},
  {Armas Padilla}, {Kaur}, {Yang}, {Soleri}, \& {Wijnands}}]{altamirano11}
{Altamirano}, D., {Cavecchi}, Y., {Patruno}, A., {et~al.} 2011, \apjl, 727,
  L18, \dodoi{10.1088/2041-8205/727/1/L18}

\bibitem[{{Applegate} \& {Shaham}(1994)}]{applegate94}
{Applegate}, J.~H., \& {Shaham}, J. 1994, \apj, 436, 312,
  \dodoi{10.1086/174906}

\bibitem[{{Arnaud}(1996)}]{arnaud96}
{Arnaud}, K.~A. 1996, in Astronomical Society of the Pacific Conference Series,
  Vol. 101, Astronomical Data Analysis Software and Systems V, ed. G.~H.
  {Jacoby} \& J.~{Barnes}, 17

\bibitem[{{Astropy Collaboration} {et~al.}(2013){Astropy Collaboration},
  {Robitaille}, {Tollerud}, {Greenfield}, {Droettboom}, {Bray}, {Aldcroft},
  {Davis}, {Ginsburg}, {Price-Whelan}, {Kerzendorf}, {Conley}, {Crighton},
  {Barbary}, {Muna}, {Ferguson}, {Grollier}, {Parikh}, {Nair}, {Unther},
  {Deil}, {Woillez}, {Conseil}, {Kramer}, {Turner}, {Singer}, {Fox}, {Weaver},
  {Zabalza}, {Edwards}, {Azalee Bostroem}, {Burke}, {Casey}, {Crawford},
  {Dencheva}, {Ely}, {Jenness}, {Labrie}, {Lim}, {Pierfederici}, {Pontzen},
  {Ptak}, {Refsdal}, {Servillat}, \& {Streicher}}]{astropy:2013}
{Astropy Collaboration}, {Robitaille}, T.~P., {Tollerud}, E.~J., {et~al.} 2013,
  \aap, 558, A33, \dodoi{10.1051/0004-6361/201322068}

\bibitem[{{Astropy Collaboration} {et~al.}(2018){Astropy Collaboration},
  {Price-Whelan}, {Sip{\H{o}}cz}, {G{\"u}nther}, {Lim}, {Crawford}, {Conseil},
  {Shupe}, {Craig}, {Dencheva}, {Ginsburg}, {Vand erPlas}, {Bradley},
  {P{\'e}rez-Su{\'a}rez}, {de Val-Borro}, {Aldcroft}, {Cruz}, {Robitaille},
  {Tollerud}, {Ardelean}, {Babej}, {Bach}, {Bachetti}, {Bakanov}, {Bamford},
  {Barentsen}, {Barmby}, {Baumbach}, {Berry}, {Biscani}, {Boquien}, {Bostroem},
  {Bouma}, {Brammer}, {Bray}, {Breytenbach}, {Buddelmeijer}, {Burke},
  {Calderone}, {Cano Rodr{\'\i}guez}, {Cara}, {Cardoso}, {Cheedella}, {Copin},
  {Corrales}, {Crichton}, {D'Avella}, {Deil}, {Depagne}, {Dietrich}, {Donath},
  {Droettboom}, {Earl}, {Erben}, {Fabbro}, {Ferreira}, {Finethy}, {Fox},
  {Garrison}, {Gibbons}, {Goldstein}, {Gommers}, {Greco}, {Greenfield},
  {Groener}, {Grollier}, {Hagen}, {Hirst}, {Homeier}, {Horton}, {Hosseinzadeh},
  {Hu}, {Hunkeler}, {Ivezi{\'c}}, {Jain}, {Jenness}, {Kanarek}, {Kendrew},
  {Kern}, {Kerzendorf}, {Khvalko}, {King}, {Kirkby}, {Kulkarni}, {Kumar},
  {Lee}, {Lenz}, {Littlefair}, {Ma}, {Macleod}, {Mastropietro}, {McCully},
  {Montagnac}, {Morris}, {Mueller}, {Mumford}, {Muna}, {Murphy}, {Nelson},
  {Nguyen}, {Ninan}, {N{\"o}the}, {Ogaz}, {Oh}, {Parejko}, {Parley}, {Pascual},
  {Patil}, {Patil}, {Plunkett}, {Prochaska}, {Rastogi}, {Reddy Janga},
  {Sabater}, {Sakurikar}, {Seifert}, {Sherbert}, {Sherwood-Taylor}, {Shih},
  {Sick}, {Silbiger}, {Singanamalla}, {Singer}, {Sladen}, {Sooley},
  {Sornarajah}, {Streicher}, {Teuben}, {Thomas}, {Tremblay}, {Turner},
  {Terr{\'o}n}, {van Kerkwijk}, {de la Vega}, {Watkins}, {Weaver}, {Whitmore},
  {Woillez}, {Zabalza}, \& {Astropy Contributors}}]{astropy:2018}
{Astropy Collaboration}, {Price-Whelan}, A.~M., {Sip{\H{o}}cz}, B.~M., {et~al.}
  2018, \aj, 156, 123, \dodoi{10.3847/1538-3881/aabc4f}

\bibitem[{{Baglio} {et~al.}(2024){Baglio}, {Russell}, {Saikia}, {Motta},
  {Mariani}, {Alabarta}, {Campana}, {Covino}, {D'Avanzo}, {Goldoni}, {Masetti},
  {Munoz-Darias}, \& {Rout}}]{atel16487}
{Baglio}, M.~C., {Russell}, D.~M., {Saikia}, P., {et~al.} 2024, The
  Astronomer's Telegram, 16487, 1

\bibitem[{{Barri{\`e}re} {et~al.}(2015){Barri{\`e}re}, {Krivonos}, {Tomsick},
  {Bachetti}, {Boggs}, {Chakrabarty}, {Christensen}, {Craig}, {Hailey},
  {Harrison}, {Hong}, {Mori}, {Stern}, \& {Zhang}}]{barriere15}
{Barri{\`e}re}, N.~M., {Krivonos}, R., {Tomsick}, J.~A., {et~al.} 2015, \apj,
  799, 123, \dodoi{10.1088/0004-637X/799/2/123}

\bibitem[{{Bhattacharyya} \& {Chakrabarty}(2017)}]{bhattacharyya17}
{Bhattacharyya}, S., \& {Chakrabarty}, D. 2017, \apj, 835, 4,
  \dodoi{10.3847/1538-4357/835/1/4}

\bibitem[{{Bostanc{\i}} {et~al.}(2023){Bostanc{\i}}, {Boztepe}, {G{\"u}ver},
  {Strohmayer}, {Cavecchi}, {G{\"o}{\u{g}}{\"u}{\c{s}}}, {Altamirano}, {Bult},
  {Chakrabarty}, {Guillot}, {Jaisawal}, {Malacaria}, {Mancuso}, {Sanna}, \&
  {Swank}}]{bostanci23}
{Bostanc{\i}}, Z.~F., {Boztepe}, T., {G{\"u}ver}, T., {et~al.} 2023, \apj, 958,
  55, \dodoi{10.3847/1538-4357/acfc4c}

\bibitem[{{Buccheri} {et~al.}(1983){Buccheri}, {Bennett}, {Bignami}, {Bloemen},
  {Boriakoff}, {Caraveo}, {Hermsen}, {Kanbach}, {Manchester}, {Masnou},
  {Mayer-Hasselwander}, {{\"O}zel}, {Paul}, {Sacco}, {Scarsi}, \&
  {Strong}}]{buccheri83}
{Buccheri}, R., {Bennett}, K., {Bignami}, G.~F., {et~al.} 1983, \aap, 128, 245

\bibitem[{{Bult} {et~al.}(2021{\natexlab{a}}){Bult}, {Strohmayer}, {Malacaria},
  {Ng}, \& {Wadiasingh}}]{bult21a}
{Bult}, P., {Strohmayer}, T.~E., {Malacaria}, C., {Ng}, M., \& {Wadiasingh}, Z.
  2021{\natexlab{a}}, \apj, 912, 120, \dodoi{10.3847/1538-4357/abf13f}

\bibitem[{{Bult} {et~al.}(2021{\natexlab{b}}){Bult}, {Altamirano},
  {Arzoumanian}, {Bilous}, {Chakrabarty}, {Gendreau}, {G{\"u}ver}, {Jaisawal},
  {Kuulkers}, {Malacaria}, {Ng}, {Sanna}, \& {Strohmayer}}]{bult21b}
{Bult}, P., {Altamirano}, D., {Arzoumanian}, Z., {et~al.} 2021{\natexlab{b}},
  \apj, 907, 79, \dodoi{10.3847/1538-4357/abd54b}

\bibitem[{{Chakrabarty} {et~al.}(2003){Chakrabarty}, {Morgan}, {Muno},
  {Galloway}, {Wijnands}, {van der Klis}, \& {Markwardt}}]{chakrabarty03}
{Chakrabarty}, D., {Morgan}, E.~H., {Muno}, M.~P., {et~al.} 2003, \nat, 424,
  42, \dodoi{10.1038/nature01732}

\bibitem[{{Chakraborty} {et~al.}(2011){Chakraborty}, {Bhattacharyya}, \&
  {Mukherjee}}]{chakraborty11}
{Chakraborty}, M., {Bhattacharyya}, S., \& {Mukherjee}, A. 2011, \mnras, 418,
  490, \dodoi{10.1111/j.1365-2966.2011.19499.x}

\bibitem[{{Chenevez} {et~al.}(2016){Chenevez}, {Galloway}, {in 't Zand},
  {Tomsick}, {Barret}, {Chakrabarty}, {F{\"u}rst}, {Boggs}, {Christensen},
  {Craig}, {Hailey}, {Harrison}, {Romano}, {Stern}, \& {Zhang}}]{chenevez16}
{Chenevez}, J., {Galloway}, D.~K., {in 't Zand}, J.~J.~M., {et~al.} 2016, \apj,
  818, 135, \dodoi{10.3847/0004-637X/818/2/135}

\bibitem[{{Cowie} {et~al.}(2024){Cowie}, {Gillanders}, {Rhodes}, {Smartt},
  {Heywood}, \& {Fender}}]{atel16477}
{Cowie}, F.~J., {Gillanders}, J.~H., {Rhodes}, L., {et~al.} 2024, The
  Astronomer's Telegram, 16477, 1

\bibitem[{da~Costa-Luis {et~al.}(2022)da~Costa-Luis, Larroque, Altendorf, Mary,
  richardsheridan, Korobov, Raphael, Ivanov, Bargull, Rodrigues, Chen, Lee,
  Newey, James, JC, Zugnoni, Pagel, mjstevens777, Dektyarev, Rothberg,
  Alexander, Panteleit, Dill, FichteFoll, Sturm, HeoHeo, van Kemenade,
  McCracken, \& MapleCCC}]{dacostaluis22}
da~Costa-Luis, C., Larroque, S.~K., Altendorf, K., {et~al.} 2022, {tqdm: A
  fast, Extensible Progress Bar for Python and CLI}, v4.64.0,  Zenodo,
  \dodoi{10.5281/zenodo.6412640}

\bibitem[{{de Jager} {et~al.}(1989){de Jager}, {Raubenheimer}, \&
  {Swanepoel}}]{dejager89}
{de Jager}, O.~C., {Raubenheimer}, B.~C., \& {Swanepoel}, J.~W.~H. 1989, \aap,
  221, 180

\bibitem[{Deloye \& Bildsten(2003)}]{deloye03}
Deloye, C.~J., \& Bildsten, L. 2003, The Astrophysical Journal, 598,
  1217–1228, \dodoi{10.1086/379063}

\bibitem[{{Di Salvo} {et~al.}(2023){Di Salvo}, {Papitto}, {Marino}, {Iaria}, \&
  {Burderi}}]{disalvo23}
{Di Salvo}, T., {Papitto}, A., {Marino}, A., {Iaria}, R., \& {Burderi}, L.
  2023, arXiv e-prints, arXiv:2311.12516, \dodoi{10.48550/arXiv.2311.12516}

\bibitem[{{Di Salvo} \& {Sanna}(2022)}]{disalvo22}
{Di Salvo}, T., \& {Sanna}, A. 2022, in Astrophysics and Space Science Library,
  Vol. 465, Astrophysics and Space Science Library, ed. S.~{Bhattacharyya},
  A.~{Papitto}, \& D.~{Bhattacharya}, 87--124,
  \dodoi{10.1007/978-3-030-85198-9_4}

\bibitem[{{Drilling} \& {Landolt}(2000)}]{drilling00}
{Drilling}, J.~S., \& {Landolt}, A.~U. 2000, in Allen's Astrophysical
  Quantities, ed. A.~N. {Cox}, 381

\bibitem[{{Folkner} {et~al.}(2009){Folkner}, {Williams}, \&
  {Boggs}}]{folkner09}
{Folkner}, W.~M., {Williams}, J.~G., \& {Boggs}, D.~H. 2009, Interplanetary
  Network Progress Report, 42-178, 1

\bibitem[{{Fujimoto} {et~al.}(1987){Fujimoto}, {Sztajno}, {Lewin}, \& {van
  Paradijs}}]{fujimoto87}
{Fujimoto}, M.~Y., {Sztajno}, M., {Lewin}, W. H.~G., \& {van Paradijs}, J.
  1987, \apj, 319, 902, \dodoi{10.1086/165507}

\bibitem[{{Galloway} {et~al.}(2004){Galloway}, {Cumming}, {Kuulkers},
  {Bildsten}, {Chakrabarty}, \& {Rothschild}}]{galloway04}
{Galloway}, D.~K., {Cumming}, A., {Kuulkers}, E., {et~al.} 2004, \apj, 601,
  466, \dodoi{10.1086/380445}

\bibitem[{{Gendreau} {et~al.}(2016){Gendreau}, {Arzoumanian}, {Adkins},
  {Albert}, {Anders}, {Aylward}, {Baker}, {Balsamo}, {Bamford}, {Benegalrao},
  {Berry}, {Bhalwani}, {Black}, {Blaurock}, {Bronke}, {Brown}, {Budinoff},
  {Cantwell}, {Cazeau}, {Chen}, {Clement}, {Colangelo}, {Coleman},
  {Coopersmith}, {Dehaven}, {Doty}, {Egan}, {Enoto}, {Fan}, {Ferro}, {Foster},
  {Galassi}, {Gallo}, {Green}, {Grosh}, {Ha}, {Hasouneh}, {Heefner}, {Hestnes},
  {Hoge}, {Jacobs}, {J{\o}rgensen}, {Kaiser}, {Kellogg}, {Kenyon}, {Koenecke},
  {Kozon}, {LaMarr}, {Lambertson}, {Larson}, {Lentine}, {Lewis}, {Lilly},
  {Liu}, {Malonis}, {Manthripragada}, {Markwardt}, {Matonak}, {Mcginnis},
  {Miller}, {Mitchell}, {Mitchell}, {Mohammed}, {Monroe}, {Montt de Garcia},
  {Mul{\'e}}, {Nagao}, {Ngo}, {Norris}, {Norwood}, {Novotka}, {Okajima},
  {Olsen}, {Onyeachu}, {Orosco}, {Peterson}, {Pevear}, {Pham}, {Pollard},
  {Pope}, {Powers}, {Powers}, {Price}, {Prigozhin}, {Ramirez}, {Reid},
  {Remillard}, {Rogstad}, {Rosecrans}, {Rowe}, {Sager}, {Sanders}, {Savadkin},
  {Saylor}, {Schaeffer}, {Schweiss}, {Semper}, {Serlemitsos}, {Shackelford},
  {Soong}, {Struebel}, {Vezie}, {Villasenor}, {Winternitz}, {Wofford},
  {Wright}, {Yang}, \& {Yu}}]{gendreau16}
{Gendreau}, K.~C., {Arzoumanian}, Z., {Adkins}, P.~W., {et~al.} 2016, in
  Society of Photo-Optical Instrumentation Engineers (SPIE) Conference Series,
  Vol. 9905, Space Telescopes and Instrumentation 2016: Ultraviolet to Gamma
  Ray, ed. J.-W.~A. {den Herder}, T.~{Takahashi}, \& M.~{Bautz}, 99051H,
  \dodoi{10.1117/12.2231304}

\bibitem[{{Ghosh} \& {Lamb}(1979)}]{ghoshlamb79}
{Ghosh}, P., \& {Lamb}, F.~K. 1979, \apj, 232, 259, \dodoi{10.1086/157285}

\bibitem[{{Girardi} {et~al.}(2000){Girardi}, {Bressan}, {Bertelli}, \&
  {Chiosi}}]{girardi00}
{Girardi}, L., {Bressan}, A., {Bertelli}, G., \& {Chiosi}, C. 2000, \aaps, 141,
  371, \dodoi{10.1051/aas:2000126}

\bibitem[{{Guiffreda} {et~al.}(2024){Guiffreda}, {De}, {Durbak}, {Kutyrev},
  {Troja}, \& {Cenk}}]{atel16499}
{Guiffreda}, O., {De}, K., {Durbak}, J., {et~al.} 2024, The Astronomer's
  Telegram, 16499, 1

\bibitem[{{G{\"u}ver} {et~al.}(2021){G{\"u}ver}, {Boztepe},
  {G{\"o}{\u{g}}{\"u}{\c{s}}}, {Chakraborty}, {Strohmayer}, {Bult},
  {Altamirano}, {Jaisawal}, {Kocab{\i}y{\i}k}, {Malacaria}, {Kashyap},
  {Gendreau}, {Arzoumanian}, \& {Chakrabarty}}]{guver21}
{G{\"u}ver}, T., {Boztepe}, T., {G{\"o}{\u{g}}{\"u}{\c{s}}}, E., {et~al.} 2021,
  \apj, 910, 37, \dodoi{10.3847/1538-4357/abe1ae}

\bibitem[{{G{\"u}ver} {et~al.}(2022{\natexlab{a}}){G{\"u}ver}, {Boztepe},
  {Ballantyne}, {Bostanc{\i}}, {Bult}, {Jaisawal}, {G{\"o}{\u{g}}{\"u}{\c{s}}},
  {Strohmayer}, {Altamirano}, {Guillot}, \& {Chakrabarty}}]{guver22a}
{G{\"u}ver}, T., {Boztepe}, T., {Ballantyne}, D.~R., {et~al.}
  2022{\natexlab{a}}, \mnras, 510, 1577, \dodoi{10.1093/mnras/stab3422}

\bibitem[{{G{\"u}ver} {et~al.}(2022{\natexlab{b}}){G{\"u}ver}, {Bostanc{\i}},
  {Boztepe}, {G{\"o}{\u{g}}{\"u}{\c{s}}}, {Bult}, {Kashyap}, {Chakraborty},
  {Ballantyne}, {Ludlam}, {Malacaria}, {Jaisawal}, {Strohmayer}, {Guillot}, \&
  {Ng}}]{guver22b}
{G{\"u}ver}, T., {Bostanc{\i}}, Z.~F., {Boztepe}, T., {et~al.}
  2022{\natexlab{b}}, \apj, 935, 154, \dodoi{10.3847/1538-4357/ac8106}

\bibitem[{{Heger} {et~al.}(2007){Heger}, {Cumming}, {Galloway}, \&
  {Woosley}}]{heger07b}
{Heger}, A., {Cumming}, A., {Galloway}, D.~K., \& {Woosley}, S.~E. 2007, \apjl,
  671, L141, \dodoi{10.1086/525522}

\bibitem[{{Hunter}(2007)}]{hunter07}
{Hunter}, J.~D. 2007, Computing in Science and Engineering, 9, 90,
  \dodoi{10.1109/MCSE.2007.55}

\bibitem[{{Illiano} {et~al.}(2023){Illiano}, {Papitto}, {Sanna}, {Bult},
  {Ambrosino}, {Miraval Zanon}, {Coti Zelati}, {Stella}, {Altamirano},
  {Baglio}, {Bozzo}, {Burderi}, {de Martino}, {Di Marco}, {di Salvo},
  {Ferrigno}, {Loktev}, {Marino}, {Ng}, {Pilia}, {Poutanen}, \&
  {Salmi}}]{illiano23}
{Illiano}, G., {Papitto}, A., {Sanna}, A., {et~al.} 2023, \apjl, 942, L40,
  \dodoi{10.3847/2041-8213/acad81}

\bibitem[{{Illiano} {et~al.}(2024){Illiano}, {Zelati}, {Marino}, {Papitto},
  {Zanon}, {Baglio}, {Del Santo}, {Russell}, \& {Ambrosino}}]{atel16510}
{Illiano}, G., {Zelati}, F.~C., {Marino}, A., {et~al.} 2024, The Astronomer's
  Telegram, 16510, 1

\bibitem[{{in't Zand} {et~al.}(2005){in't Zand}, {Cornelisse}, \&
  {M{\'e}ndez}}]{intzand05}
{in't Zand}, J.~J.~M., {Cornelisse}, R., \& {M{\'e}ndez}, M. 2005, \aap, 440,
  287, \dodoi{10.1051/0004-6361:20052955}

\bibitem[{{Kaastra} \& {Bleeker}(2016)}]{kaastra16}
{Kaastra}, J.~S., \& {Bleeker}, J.~A.~M. 2016, \aap, 587, A151,
  \dodoi{10.1051/0004-6361/201527395}

\bibitem[{{Knigge} {et~al.}(2011){Knigge}, {Baraffe}, \&
  {Patterson}}]{knigge11}
{Knigge}, C., {Baraffe}, I., \& {Patterson}, J. 2011, \apjs, 194, 28,
  \dodoi{10.1088/0067-0049/194/2/28}

\bibitem[{{Kuulkers} {et~al.}(2003){Kuulkers}, {den Hartog}, {in't Zand},
  {Verbunt}, {Harris}, \& {Cocchi}}]{kuulkers03}
{Kuulkers}, E., {den Hartog}, P.~R., {in't Zand}, J.~J.~M., {et~al.} 2003,
  \aap, 399, 663, \dodoi{10.1051/0004-6361:20021781}

\bibitem[{{LaMarr} {et~al.}(2016){LaMarr}, {Prigozhin}, {Remillard}, {Malonis},
  {Gendreau}, {Arzoumanian}, {Markwardt}, \& {Baumgartner}}]{lamarr16}
{LaMarr}, B., {Prigozhin}, G., {Remillard}, R., {et~al.} 2016, in Society of
  Photo-Optical Instrumentation Engineers (SPIE) Conference Series, Vol. 9905,
  Space Telescopes and Instrumentation 2016: Ultraviolet to Gamma Ray, ed.
  J.-W.~A. {den Herder}, T.~{Takahashi}, \& M.~{Bautz}, 99054W,
  \dodoi{10.1117/12.2232784}

\bibitem[{{Lampe} {et~al.}(2016){Lampe}, {Heger}, \& {Galloway}}]{lampe16}
{Lampe}, N., {Heger}, A., \& {Galloway}, D.~K. 2016, \apj, 819, 46,
  \dodoi{10.3847/0004-637X/819/1/46}

\bibitem[{{Lange} {et~al.}(2001){Lange}, {Camilo}, {Wex}, {Kramer}, {Backer},
  {Lyne}, \& {Doroshenko}}]{lange01}
{Lange}, C., {Camilo}, F., {Wex}, N., {et~al.} 2001, \mnras, 326, 274,
  \dodoi{10.1046/j.1365-8711.2001.04606.x}

\bibitem[{{Lasota}(2001)}]{lasota01}
{Lasota}, J.-P. 2001, \nar, 45, 449, \dodoi{10.1016/S1387-6473(01)00112-9}

\bibitem[{{Li} {et~al.}(2024){Li}, {Kuiper}, {Falanga}, {Xu}, {Pan}, {Zhang},
  {Zhang}, {Qu}, {Song}, {Chen}, {Jia}, {Li}, {Zheng}, {Huang}, \&
  {Poutanen}}]{atel16548}
{Li}, Z., {Kuiper}, L., {Falanga}, M., {et~al.} 2024, The Astronomer's
  Telegram, 16548, 1

\bibitem[{{Liddle}(2007)}]{liddle07}
{Liddle}, A.~R. 2007, \mnras, 377, L74,
  \dodoi{10.1111/j.1745-3933.2007.00306.x}

\bibitem[{{Luo} {et~al.}(2021){Luo}, {Ransom}, {Demorest}, {Ray}, {Archibald},
  {Kerr}, {Jennings}, {Bachetti}, {van Haasteren}, {Champagne}, {Colen},
  {Phillips}, {Zimmerman}, {Stovall}, {Lam}, \& {Jenet}}]{luo21}
{Luo}, J., {Ransom}, S., {Demorest}, P., {et~al.} 2021, \apj, 911, 45,
  \dodoi{10.3847/1538-4357/abe62f}

\bibitem[{{Mariani} {et~al.}(2024){Mariani}, {Motta}, {Baglio}, {Fender}, {van
  den Eijden}, {X-KAT Collaboration}, {Mereminskiy}, \&
  {Lutovinov}}]{atel16475}
{Mariani}, I., {Motta}, S., {Baglio}, M.~C., {et~al.} 2024, The Astronomer's
  Telegram, 16475, 1

\bibitem[{{Marino} {et~al.}(2019{\natexlab{a}}){Marino}, {Di Salvo}, {Burderi},
  {Sanna}, {Riggio}, {Papitto}, {Del Santo}, {Gambino}, {Iaria}, \&
  {Mazzola}}]{marino19a}
{Marino}, A., {Di Salvo}, T., {Burderi}, L., {et~al.} 2019{\natexlab{a}}, \aap,
  627, A125, \dodoi{10.1051/0004-6361/201834460}

\bibitem[{{Marino} {et~al.}(2019{\natexlab{b}}){Marino}, {Del Santo}, {Cocchi},
  {D'A{\i}}, {Segreto}, {Ferrigno}, {Di Salvo}, {Malzac}, {Iaria}, \&
  {Burderi}}]{marino19b}
{Marino}, A., {Del Santo}, M., {Cocchi}, M., {et~al.} 2019{\natexlab{b}},
  \mnras, 490, 2300, \dodoi{10.1093/mnras/stz2726}

\bibitem[{{Meisel}(2018)}]{meisel18}
{Meisel}, Z. 2018, \apj, 860, 147, \dodoi{10.3847/1538-4357/aac3d3}

\bibitem[{{Mereminskiy} {et~al.}(2024){Mereminskiy}, {Semena}, {Molkov},
  {Lutovinov}, {Tkachenko}, \& {Arefiev}}]{atel16464}
{Mereminskiy}, I.~A., {Semena}, A.~N., {Molkov}, S.~V., {et~al.} 2024, The
  Astronomer's Telegram, 16464, 1

\bibitem[{{Mihara} {et~al.}(2024){Mihara}, {Negoro}, {Nakajima}, {Nagashima},
  {Kobayashi}, {Tanaka}, {Soejima}, {Kudo}, {Kawamuro}, {Yamada}, {Wang},
  {Tamagawa}, {Kawai}, {Matsuoka}, {Sakamoto}, {Serino}, {Sugita}, {Hiramatsu},
  {Nishikawa}, {Yoshida}, {Tsuboi}, {Urabe}, {Nawa}, {Nemoto}, {Goto},
  {Shidatsu}, {Takahashi}, {Niwano}, {Sato}, {Higuchi}, {Yatsu}, {Nakahira},
  {Ueno}, {Tomida}, {Ishikawa}, {Ogawa}, {Kurihara}, {Ueda}, {Setoguchi},
  {Yoshitake}, {Nakatani}, {Okada}, {Yamauchi}, {Hagiwara}, {Umeki}, {Otsuki},
  {Yamaoka}, {Kawakubo}, {Sugizaki}, \& {Iwakiri}}]{atel16469}
{Mihara}, T., {Negoro}, H., {Nakajima}, M., {et~al.} 2024, The Astronomer's
  Telegram, 16469, 1

\bibitem[{{Molkov} {et~al.}(2024){Molkov}, {Lutovinov}, {Tsygankov},
  {Suleimanov}, {Poutanen}, {Lapshov}, {Mereminskiy}, {Semena}, {Arefiev}, \&
  {Tkachenko}}]{molkov24}
{Molkov}, S.~V., {Lutovinov}, A.~A., {Tsygankov}, S.~S., {et~al.} 2024, arXiv
  e-prints, arXiv:2404.19709, \dodoi{10.48550/arXiv.2404.19709}

\bibitem[{{Motta} {et~al.}(2011){Motta}, {D'A{\`\i}}, {Papitto}, {Riggio}, {di
  Salvo}, {Burderi}, {Belloni}, {Stella}, \& {Iaria}}]{papitto11a}
{Motta}, S., {D'A{\`\i}}, A., {Papitto}, A., {et~al.} 2011, \mnras, 414, 1508,
  \dodoi{10.1111/j.1365-2966.2011.18483.x}

\bibitem[{{Mukherjee} {et~al.}(2015){Mukherjee}, {Bult}, {van der Klis}, \&
  {Bhattacharya}}]{mukherjee15}
{Mukherjee}, D., {Bult}, P., {van der Klis}, M., \& {Bhattacharya}, D. 2015,
  \mnras, 452, 3994, \dodoi{10.1093/mnras/stv1542}

\bibitem[{{NASA High Energy Astrophysics Science Archive Research Center
  (HEASARC)}(2014)}]{heasoft}
{NASA High Energy Astrophysics Science Archive Research Center (HEASARC)}.
  2014, {HEAsoft: Unified Release of FTOOLS and XANADU}, Astrophysics Source
  Code Library, record ascl:1408.004.
\newblock \doeprint{1408.004}

\bibitem[{{Negoro} {et~al.}(2024){Negoro}, {Mihara}, {Serino}, {Nakajima},
  {Kobayashi}, {Tanaka}, {Soejima}, {Kudo}, {Kawamuro}, {Yamada}, {Wang},
  {Tamagawa}, {Kawai}, {Matsuoka}, {Sakamoto}, {Sugita}, {Hiramatsu},
  {Nishikawa}, {Yoshida}, {Tsuboi}, {Urabe}, {Nawa}, {Nemoto}, {Goto},
  {Shidatsu}, {Niida}, {Takahashi}, {Niwano}, {Sato}, {Higuchi}, {Yatsu},
  {Nakahira}, {Ueno}, {Tomida}, {Ishikawa}, {Ogawa}, {Kurihara}, {Ueda},
  {Setoguchi}, {Yoshitake}, {Nakatani}, {Okada}, {Yamauchi}, {Hagiwara},
  {Umeki}, {Otsuki}, {Yamaoka}, {Kawakubo}, {Sugizaki}, \&
  {Iwakiri}}]{atel16483}
{Negoro}, H., {Mihara}, T., {Serino}, M., {et~al.} 2024, The Astronomer's
  Telegram, 16483, 1

\bibitem[{{Ng} {et~al.}(2021){Ng}, {Ray}, {Bult}, {Chakrabarty}, {Jaisawal},
  {Malacaria}, {Altamirano}, {Arzoumanian}, {Gendreau}, {G{\"u}ver}, {Kerr},
  {Strohmayer}, {Wadiasingh}, \& {Wolff}}]{ngmason21}
{Ng}, M., {Ray}, P.~S., {Bult}, P., {et~al.} 2021, \apjl, 908, L15,
  \dodoi{10.3847/2041-8213/abe1b4}

\bibitem[{{Ng} {et~al.}(2024){Ng}, {Sanna}, {Strohmayer}, {Arzoumanian},
  {Gendreau}, {Ray}, {Coley}, {Chakrabarty}, {Guillot}, {Bogdanov},
  {Altamirano}, {Chenevez}, {Hare}, {Wolff}, {Guver}, {Jaisawal}, {Wadiasingh},
  \& {Ferrara}}]{atel16474}
{Ng}, M., {Sanna}, A., {Strohmayer}, T.~E., {et~al.} 2024, The Astronomer's
  Telegram, 16474, 1

\bibitem[{{Papitto} {et~al.}(2007){Papitto}, {di Salvo}, {Burderi}, {Menna},
  {Lavagetto}, \& {Riggio}}]{papitto07}
{Papitto}, A., {di Salvo}, T., {Burderi}, L., {et~al.} 2007, \mnras, 375, 971,
  \dodoi{10.1111/j.1365-2966.2006.11359.x}

\bibitem[{{Papitto} {et~al.}(2011){Papitto}, {Bozzo}, {Ferrigno}, {Belloni},
  {Burderi}, {di Salvo}, {Riggio}, {D'A{\`\i}}, \& {Iaria}}]{papitto11b}
{Papitto}, A., {Bozzo}, E., {Ferrigno}, C., {et~al.} 2011, \aap, 535, L4,
  \dodoi{10.1051/0004-6361/201117995}

\bibitem[{{Papitto} {et~al.}(2013){Papitto}, {Ferrigno}, {Bozzo}, {Rea},
  {Pavan}, {Burderi}, {Burgay}, {Campana}, {di Salvo}, {Falanga},
  {Filipovi{\'c}}, {Freire}, {Hessels}, {Possenti}, {Ransom}, {Riggio},
  {Romano}, {Sarkissian}, {Stairs}, {Stella}, {Torres}, {Wieringa}, \&
  {Wong}}]{papitto13}
{Papitto}, A., {Ferrigno}, C., {Bozzo}, E., {et~al.} 2013, \nat, 501, 517,
  \dodoi{10.1038/nature12470}

\bibitem[{{Patruno} \& {Watts}(2021)}]{patruno21}
{Patruno}, A., \& {Watts}, A.~L. 2021, in Astrophysics and Space Science
  Library, Vol. 461, Timing Neutron Stars: Pulsations, Oscillations and
  Explosions, ed. T.~M. {Belloni}, M.~{M{\'e}ndez}, \& C.~{Zhang}, 143--208,
  \dodoi{10.1007/978-3-662-62110-3_4}

\bibitem[{{Pavlinsky} {et~al.}(2021){Pavlinsky}, {Tkachenko}, {Levin},
  {Alexandrovich}, {Arefiev}, {Babyshkin}, {Batanov}, {Bodnar}, {Bogomolov},
  {Bubnov}, {Buntov}, {Burenin}, {Chelovekov}, {Chen}, {Drozdova}, {Ehlert},
  {Filippova}, {Frolov}, {Gamkov}, {Garanin}, {Garin}, {Glushenko}, {Gorelov},
  {Grebenev}, {Grigorovich}, {Gureev}, {Gurova}, {Ilkaev}, {Katasonov},
  {Krivchenko}, {Krivonos}, {Korotkov}, {Kudelin}, {Kuznetsova}, {Lazarchuk},
  {Lomakin}, {Lapshov}, {Lipilin}, {Lutovinov}, {Mereminskiy}, {Molkov},
  {Nazarov}, {Oleinikov}, {Pikalov}, {Ramsey}, {Roiz}, {Rotin}, {Ryadov},
  {Sankin}, {Sazonov}, {Sedov}, {Semena}, {Semena}, {Serbinov}, {Shirshakov},
  {Shtykovsky}, {Shvetsov}, {Sunyaev}, {Swartz}, {Tambov}, {Voron}, \&
  {Yaskovich}}]{pavlinsky21}
{Pavlinsky}, M., {Tkachenko}, A., {Levin}, V., {et~al.} 2021, \aap, 650, A42,
  \dodoi{10.1051/0004-6361/202040265}

\bibitem[{{Perez} \& {Granger}(2007)}]{perez07}
{Perez}, F., \& {Granger}, B.~E. 2007, Computing in Science and Engineering, 9,
  21, \dodoi{10.1109/MCSE.2007.53}

\bibitem[{{Prigozhin} {et~al.}(2016){Prigozhin}, {Gendreau}, {Doty}, {Foster},
  {Remillard}, {Malonis}, {LaMarr}, {Vezie}, {Egan}, {Villasenor},
  {Arzoumanian}, {Baumgartner}, {Scholze}, {Laubis}, {Krumrey}, \&
  {Huber}}]{prigozhin16}
{Prigozhin}, G., {Gendreau}, K., {Doty}, J.~P., {et~al.} 2016, in Society of
  Photo-Optical Instrumentation Engineers (SPIE) Conference Series, Vol. 9905,
  Space Telescopes and Instrumentation 2016: Ultraviolet to Gamma Ray, ed.
  J.-W.~A. {den Herder}, T.~{Takahashi}, \& M.~{Bautz}, 99051I,
  \dodoi{10.1117/12.2231718}

\bibitem[{{Ransom}(2011)}]{ransom11}
{Ransom}, S. 2011, {PRESTO: PulsaR Exploration and Search TOolkit},
  Astrophysics Source Code Library, record ascl:1107.017

\bibitem[{{Ransom} {et~al.}(2002){Ransom}, {Eikenberry}, \&
  {Middleditch}}]{ransom02}
{Ransom}, S.~M., {Eikenberry}, S.~S., \& {Middleditch}, J. 2002, \aj, 124,
  1788, \dodoi{10.1086/342285}

\bibitem[{{Ray} {et~al.}(2024){Ray}, {Strohmayer}, {Sanna}, {Ng},
  {Arzoumanian}, {Gendreau}, {Ferrara}, {Coley}, \& {Bogdanov}}]{atel16480}
{Ray}, P.~S., {Strohmayer}, T.~E., {Sanna}, A., {et~al.} 2024, The Astronomer's
  Telegram, 16480, 1

\bibitem[{{Remillard} {et~al.}(2022){Remillard}, {Loewenstein}, {Steiner},
  {Prigozhin}, {LaMarr}, {Enoto}, {Gendreau}, {Arzoumanian}, {Markwardt},
  {Basak}, {Stevens}, {Ray}, {Altamirano}, \& {Buisson}}]{remillard22}
{Remillard}, R.~A., {Loewenstein}, M., {Steiner}, J.~F., {et~al.} 2022, \aj,
  163, 130, \dodoi{10.3847/1538-3881/ac4ae6}

\bibitem[{{Riggio} {et~al.}(2011){Riggio}, {Papitto}, {Burderi}, {di Salvo},
  {Bachetti}, {Iaria}, {D'A{\`\i}}, \& {Menna}}]{riggio11}
{Riggio}, A., {Papitto}, A., {Burderi}, L., {et~al.} 2011, \aap, 526, A95,
  \dodoi{10.1051/0004-6361/201014322}

\bibitem[{{Romani}(1990)}]{romani90}
{Romani}, R.~W. 1990, \nat, 347, 741, \dodoi{10.1038/347741a0}

\bibitem[{{Romani} {et~al.}(2022){Romani}, {Kandel}, {Filippenko}, {Brink}, \&
  {Zheng}}]{romani22}
{Romani}, R.~W., {Kandel}, D., {Filippenko}, A.~V., {Brink}, T.~G., \& {Zheng},
  W. 2022, \apjl, 934, L17, \dodoi{10.3847/2041-8213/ac8007}

\bibitem[{{Russell} {et~al.}(2024){Russell}, {Carotenuto}, {Eijnden},
  {Kuulkers}, {Sanchez-Fernandez}, {Degenaar}, {Fijma}, {Maccarone},
  {Miller-Jones}, {Tetarenko}, {Marino}, {Del Santo}, \&
  {Gusinskaia}}]{atel16511}
{Russell}, T.~D., {Carotenuto}, F., {Eijnden}, J. v.~d., {et~al.} 2024, The
  Astronomer's Telegram, 16511, 1

\bibitem[{{Saikia} {et~al.}(2024){Saikia}, {Russell}, {Baglio}, {Alabarta},
  {Rout}, \& {Lewis}}]{atel16489}
{Saikia}, P., {Russell}, D.~M., {Baglio}, M.~C., {et~al.} 2024, The
  Astronomer's Telegram, 16489, 1

\bibitem[{{Sanchez-Fernandez} {et~al.}(2024{\natexlab{a}}){Sanchez-Fernandez},
  {Kuulkers}, {Ferrigno}, \& {Chenevez}}]{atel16485}
{Sanchez-Fernandez}, C., {Kuulkers}, E., {Ferrigno}, C., \& {Chenevez}, J.
  2024{\natexlab{a}}, The Astronomer's Telegram, 16485, 1

\bibitem[{{Sanchez-Fernandez} {et~al.}(2024{\natexlab{b}}){Sanchez-Fernandez},
  {Kuulkers}, {Ferrigno}, {Chenevez}, \& {Del Santo}}]{atel16507}
{Sanchez-Fernandez}, C., {Kuulkers}, E., {Ferrigno}, C., {Chenevez}, J., \&
  {Del Santo}, M. 2024{\natexlab{b}}, The Astronomer's Telegram, 16507, 1

\bibitem[{{Sanna} {et~al.}(2018){Sanna}, {Ferrigno}, {Ray}, {Ducci},
  {Jaisawal}, {Enoto}, {Bozzo}, {Altamirano}, {Di Salvo}, {Strohmayer},
  {Papitto}, {Riggio}, {Burderi}, {Bult}, {Bogdanov}, {Gambino}, {Marino},
  {Iaria}, {Arzoumanian}, {Chakrabarty}, {Gendreau}, {Guillot}, {Markwardt}, \&
  {Wolff}}]{sanna18}
{Sanna}, A., {Ferrigno}, C., {Ray}, P.~S., {et~al.} 2018, \aap, 617, L8,
  \dodoi{10.1051/0004-6361/201834160}

\bibitem[{{Sguera} \& {Sidoli}(2024)}]{atel16493}
{Sguera}, V., \& {Sidoli}, L. 2024, The Astronomer's Telegram, 16493, 1

\bibitem[{{Sokolovsky} {et~al.}(2024){Sokolovsky}, {Korotkiy}, \&
  {Zalles}}]{atel16476}
{Sokolovsky}, K., {Korotkiy}, S., \& {Zalles}, R. 2024, The Astronomer's
  Telegram, 16476, 1

\bibitem[{{Strohmayer} {et~al.}(2010){Strohmayer}, {Markwardt}, {Pereira}, \&
  {Smith}}]{strohmayer10}
{Strohmayer}, T.~E., {Markwardt}, C.~B., {Pereira}, D., \& {Smith}, E.~A. 2010,
  The Astronomer's Telegram, 2946, 1

\bibitem[{{Sunyaev} {et~al.}(2021){Sunyaev}, {Arefiev}, {Babyshkin},
  {Bogomolov}, {Borisov}, {Buntov}, {Brunner}, {Burenin}, {Churazov},
  {Coutinho}, {Eder}, {Eismont}, {Freyberg}, {Gilfanov}, {Gureyev}, {Hasinger},
  {Khabibullin}, {Kolmykov}, {Komovkin}, {Krivonos}, {Lapshov}, {Levin},
  {Lomakin}, {Lutovinov}, {Medvedev}, {Merloni}, {Mernik}, {Mikhailov},
  {Molodtsov}, {Mzhelsky}, {M{\"u}ller}, {Nandra}, {Nazarov}, {Pavlinsky},
  {Poghodin}, {Predehl}, {Robrade}, {Sazonov}, {Scheuerle}, {Shirshakov},
  {Tkachenko}, \& {Voron}}]{sunyaev21}
{Sunyaev}, R., {Arefiev}, V., {Babyshkin}, V., {et~al.} 2021, \aap, 656, A132,
  \dodoi{10.1051/0004-6361/202141179}

\bibitem[{{Takeda} {et~al.}(2024){Takeda}, {Ota}, {Watanabe}, {Aoyama},
  {Iwata}, {Tamagawa}, {Enoto}, {Kitaguchi}, {Iwakiri}, {Kato}, {Hu}, {Mihara},
  \& {NinjaSat Team}}]{atel16495}
{Takeda}, T., {Ota}, N., {Watanabe}, S., {et~al.} 2024, The Astronomer's
  Telegram, 16495, 1

\bibitem[{{Tout} {et~al.}(1996){Tout}, {Pols}, {Eggleton}, \& {Han}}]{tout96}
{Tout}, C.~A., {Pols}, O.~R., {Eggleton}, P.~P., \& {Han}, Z. 1996, \mnras,
  281, 257, \dodoi{10.1093/mnras/281.1.257}

\bibitem[{{Ubertini} {et~al.}(1999){Ubertini}, {Bazzano}, {Cocchi},
  {Natalucci}, {Heise}, {Muller}, \& {in 't Zand}}]{ubertini99}
{Ubertini}, P., {Bazzano}, A., {Cocchi}, M., {et~al.} 1999, \apjl, 514, L27,
  \dodoi{10.1086/311933}

\bibitem[{Virtanen {et~al.}(2020)Virtanen, Gommers, Oliphant, Haberland, Reddy,
  Cournapeau, Burovski, Peterson, Weckesser, Bright, {van der Walt}, Brett,
  Wilson, Millman, Mayorov, Nelson, Jones, Kern, Larson, Carey, Polat, Feng,
  Moore, {VanderPlas}, Laxalde, Perktold, Cimrman, Henriksen, Quintero, Harris,
  Archibald, Ribeiro, Pedregosa, {van Mulbregt}, \& {SciPy 1.0
  Contributors}}]{virtanen20}
Virtanen, P., Gommers, R., Oliphant, T.~E., {et~al.} 2020, Nature Methods, 17,
  261, \dodoi{10.1038/s41592-019-0686-2}

\bibitem[{{Wilms} {et~al.}(2000){Wilms}, {Allen}, \& {McCray}}]{wilms00}
{Wilms}, J., {Allen}, A., \& {McCray}, R. 2000, \apj, 542, 914,
  \dodoi{10.1086/317016}

\bibitem[{{Worpel} {et~al.}(2013){Worpel}, {Galloway}, \& {Price}}]{worpel13}
{Worpel}, H., {Galloway}, D.~K., \& {Price}, D.~J. 2013, \apj, 772, 94,
  \dodoi{10.1088/0004-637X/772/2/94}

\bibitem[{{Worpel} {et~al.}(2015){Worpel}, {Galloway}, \& {Price}}]{worpel15}
---. 2015, \apj, 801, 60, \dodoi{10.1088/0004-637X/801/1/60}

\bibitem[{{Yun} {et~al.}(2023){Yun}, {Grefenstette}, {Ludlam}, {Brumback},
  {Buisson}, {Mastroserio}, \& {Pike}}]{yun23}
{Yun}, S.~B., {Grefenstette}, B.~W., {Ludlam}, R.~M., {et~al.} 2023, \apj, 947,
  81, \dodoi{10.3847/1538-4357/acb689}

\end{thebibliography}
\bibliographystyle{aasjournal}

\end{document}